\documentclass{iopart}
\usepackage{graphicx}

\def\bn{\begin{itemize}}
\def\en{\end{itemize}}

\begin{document}
\bibliographystyle{unsrt}

\title{Evolution of displacements and strains in sheared amorphous solids}
\author{Craig E.~Maloney$^{(1,2)}$}
\ead{craigmaloney@cmu.edu}
\author{Mark O.~Robbins$^{(2)}$}
\ead{mr@pha.jhu.edu}

\address{$^{(1)}$ Dept. of Civil and Environmental Engineering, Carnegie Mellon University, Pittsburgh, PA, USA}
\address{$^{(2)}$Dept. of Physics and Astronomy, The Johns Hopkins University, Baltimore, MD, USA}


\begin{abstract}
The local deformation of two-dimensional Lennard-Jones glasses under imposed shear strain is studied via computer simulations.
Both the mean squared displacement and mean squared strain rise linearly with the length of the strain interval $\Delta \gamma$ over which they are measured.
However, the increase in displacement does not represent single-particle diffusion.
There are long-range spatial correlations in displacement associated with slip lines with an amplitude of order the particle size.
Strong dependence on system size is also observed.
The probability distributions of displacement and strain are very different.
For small $\Delta \gamma$ the distribution of displacement has a plateau followed by an exponential tail.
The distribution becomes Gaussian as $\Delta \gamma$ increases to about .03.
The strain distributions consist of sharp central peaks associated with elastic regions, and long exponential tails associated with plastic regions.
The latter persist to the largest $\Delta \gamma$ studied.

\end{abstract}
\maketitle

\section{Introduction}
The mechanical properties of amorphous materials are important in many structural applications and the microscopic mechanisms underlying plastic deformation have been the subject of great theoretical interest~\cite{SPAEPEN1977hc,Falk1998pd,ARGON1979vn} .
While deformation of crystalline materials can be understood through the generation and propagation of dislocations, no analogous defect has been identified in disordered materials.
This has a direct impact on their mechanical properties since
dislocation motion reduces the yield stress of crystals by orders of magnitude.
In the last decade, it has also been realized that amorphous materials share many features with colloids, granular media and other systems where motion is jammed or arrested by packing constraints \cite{Liu1998dp,Kob1997lr,Weeks2000fk}.
Common features are observed in the onset of motion under external forces in such systems and understanding their behavior represents a fundamental problem in nonequilibrium statistical mechanics.

Recent simulations and experiments have followed the motion of individual atoms or particles to understand how disordered solids deform under shear 
\cite{Falk1998pd,Kob1997lr,Weeks2000fk,ARGON1979fj,KOBAYASHI1980qy,MAEDA1981uq,SROLOVITZ1981fk,DENG1989bh,Lauridsen2002zr,Shi2005la,Ono2003fk,Ono2002qy,utterPhysRevE69031308,utter2007,PhysRevE68011507,Rottler2005uq,Varnik2004qy,Schall2007,Tanguy2006ul,Lacks2004rw,Lacks2002lr,Malandro1998pb,Lemaitre2006pb,Maloney2004dk,Maloney2004fu,Maloney2006dz,Maloney2006kl}.
One of the most studied quantities has been the van Hove function $P(\Delta r;\Delta t)$, which describes the probability distribution of the displacements $\Delta r$ of atoms as a function of time interval $\Delta t$.
Here $\Delta r$ is defined relative to a uniformly (affinely) sheared system.
In equilibrium systems, $P(\Delta r)$ has been used to determine the time scale on which particles escape from local cages \cite{Kob1997lr,Weeks2000fk,Kremmer1987} and its long time behavior gives the diffusion constant.
Studies of sheared systems have examined the changes in $P(\Delta r)$ to determine the effect of strain rate on caging times and diffusion\cite{Ono2003fk,Ono2002qy,Tanguy2006ul,Lacks2002lr,Lemaitre2007ij}.
These studies have generally found $\langle \Delta r^2 \rangle \propto \Delta t$ as expected for single-particle diffusion.
In some cases the associated nonequilibrium diffusion constant obeys a generalized fluctuation dissipation theorem with an effective temperature \cite{Ono2003fk,Ono2002qy,Lacks2002lr}.

In this work, we use molecular dynamics (MD) simulations to examine the spatio-temporal correlations in $\Delta r$ in athermal systems that are driven by steady shear.
This models the low temperature behavior of amorphous atomic systems as well as disordered granular or colloidal systems where thermal fluctuations are small.
As in most studies of similar systems \cite{Tanguy2006ul,Lacks2002lr,Lemaitre2007ij}, we find that the long time evolution of $P(\Delta r; \Delta t)$ is consistent with the diffusion equation.
However, the displacement shows strong spatial correlations that span the entire system.
Thus $P$ does not describe single-particle diffusion and the diffusion constant has a strong dependence on system size \cite{Lemaitre2007ij}.
The spatial correlations in $\Delta r$ are associated with plastic deformation along slip lines that grow to span the system.
The slip accommodated in these plastic zones is one or two molecular diameters.
This sets a characteristic strain interval between their formation that decreases with system size.
Spatio temporal correlations in the location of the slip lines are studied through images and the evolution in the functional form of $P$ with strain.

The long-range spatial correlations in $\Delta r$ reflect the fact that large regions deform elastically.
To measure the local magnitude of non-affine deformations, we evaluate the vorticity in the displacement field $\omega$.
Other measures of the \emph{relative} particle displacement have been considered in previous work, but $\omega$ has the advantage that it allows both the magnitude and direction of the displacements to be quantified.
This allows strong directional, spatial and temporal correlations in strain to be identified.

The distributions of $\Delta \vec{r}$ and $\omega$ provide additional information about correlations.
Over short intervals, both distributions show exponential tails covering up to eight orders of magnitude.
The strain characterizing the exponential decay for $\omega$ is close to the
the macroscopic yield strain of the material.
As the interval of time or strain increases, $P(\Delta \vec{r})$ becomes more isotropic and converges to a Gaussian at strain intervals of order 3\%.
The distribution of $\omega$ retains an exponential tail out to much larger strains, indicating that particles are still not undergoing single-particle diffusion.

\section{Protocol and definitions}
We perform classical 2D molecular dynamics simulations of a bi-disperse mixture that is chosen to inhibit crystallization~\cite{Maloney2006dz}.
There are $N_L$ large (L) particles and $N_S$ small (S) particles with $N_L/N_S=(1+\sqrt{5})/4$.
The interaction between particles of type $I$ and $J$ 
is a Lennard Jones (LJ) potential: $U_{IJ} = 4\epsilon\left[(r/\sigma_{IJ})^{-6}-(r/\sigma_{IJ})^{-12}\right]$
where $\epsilon$ sets the characteristic energy scale and $\sigma_{IJ}$ reflects the size of $I$ and $J$.
We set $\sigma_{LL}=1.0\sigma_0$, $\sigma_{SS}=0.6\sigma_0$, and $\sigma_{LS}=0.5 (\sigma_{LL}+\sigma_{SS}) = 0.8\sigma_0$ and measure all lengths in units of $\sigma_0$.
To speed calculations, the potential is truncated smoothly using a fourth order polynomial that matches the LJ potential and force at $r_{\rm{in}} = 1.2 \sigma_{IJ}$ and takes both to zero at $r_{\rm{out}}=1.5 \sigma_{IJ}$.
All particles have mass $m$ and the characteristic time is $\tau \equiv \sqrt{m\sigma_0^2/\epsilon}$.

The equations of motion are integrated using the parallel MD code LAMMPS \cite{lammps} with time step $\delta t = 0.0056 \tau$.
The system temperature is maintained at zero by applying a locally Galilean-invariant Kelvin damping mechanism \cite{DPD}.
Neighboring particles exert equal and opposite drag forces on each other which are proportional to their relative velocity. 
The value of the damping constant was set to $0.45 m/\tau$.
With this choice, modes with wavelengths at the particle scale are nearly critically damped and damping decreases at longer lengths.
The limit of zero temperature is of fundamental interest in atomic systems and thermal motion can generally be ignored in systems where the particles represent larger entities such as grains or colloids.

An amorphous state of the system is prepared following the protocol in previous studies~\cite{Maloney2006kl}.
Periodic boundary conditions with periods $L_x$ and $L_y$ are used.
A random configuration with a particle area fraction of $0.85$ is equilibrated for $20,000$ time steps using a soft repulsive potential to ensure particles do not overlap.
The interaction is then switched to the LJ potential, and the damped system is run for $10,000$ timesteps with fixed area.
The system is next allowed to expand toward a zero pressure state under an isotropic Nose-Hoover barostat~\cite{lammps} with a time constant of $220\tau$ 
and a damping parameter of $45~\tau^{-1}$ for $10,000$ timesteps.
A final quench is run for $10,000$ timesteps with a fixed cell size prior to shearing.

Equilibrated systems are compressed along the $y$ direction at constant true strain rate $\dot{\gamma}$ and expanded along $x$ to maintain constant area:  $L_x=L_{x0}e^{+\dot{\gamma}t}$, $L_y=L_{y0}e^{-\dot{\gamma}t}$. 
Unless noted, the results below are for about $1.6 \times10^{6}$ particles with $L_{x0}=L_{y0}\approx 1000\sigma_0,$ and $\dot{\gamma}=0.45 \times 10^{-5} \tau^{-1}$. 
Studies showed that this strain rate was slow enough to be in a quasistatic regime where results are rate-independent.
For this reason, results will be expressed in terms of the interval of strain $\Delta \gamma$ over which they were obtained, rather than in terms of time interval.
For small strains the rate of plastic deformation increases with strain and may be sensitive to initial conditions~\cite{PhysRevE68011507,Varnik2004qy,RottlerRobbins05}.
For strains greater than 6\% the behavior becomes independent of strain
and initial state. 
Data presented below are all from this steady state regime.

The non-affine component $\tilde{\bf{v}}$ of a particle's velocity $\bf{v}$ is the deviation from the velocity associated with a uniform strain:
\begin{equation}\begin{array}{c}
\tilde{v}_x = v_x - \dot{\gamma} x(\tau)\\ 
\tilde{v}_y = v_y + \dot{\gamma} y(\tau) 
\end{array} \label{eq:background} \end{equation}
Integrating over time gives the net non-affine displacement $\Delta \vec{r} = (\Delta x, \Delta y)$ over any finite strain interval.
Note that this is not the same as taking the difference between the final position of a particle and the position corresponding to an affine displacement.
The reason is that a small non-affine displacement at the beginning of a time interval is amplified by subsequent affine flow and would have the same impact as a much larger non-affine motion at the end of the interval.
A similar effect produces an extra Taylor diffusion in sheared fluids~\cite{taylorDiffusion}.
From now on we shall refer to the non-affine displacement simply as "displacement" and use "total displacement" when the affine component is not subtracted.
Note that the non-affine displacement averaged over the periodic cell is zero.
This follows because the equations of motion in the deforming cell \cite{lammps} conserve the sum over particles of the peculiar momentum $m \bf{\tilde{v}}$ and all particles have the same mass.

The results presented below show that the displacement field has long-range spatial correlations.
A more local measure of deformation is provided by the spatial derivatives of ${\Delta \bf r}$, which are obtained in the following way.
A Delaunnay triangulation is found for an initial reference configuration,
with vertices at the center of all particles.
The derivatives $\partial_i \Delta r_j$ on each triangle are then calculated from the finite difference between displacement vectors at the three vertices.
To make plots and evaluate probability distributions, the derivatives are sampled on a fine, rectilinear grid with spacing $\sim \sigma_0/3$.
An important property of this piece-wise constant definition of $\partial_i \Delta r_j$ is that its integral along an arbitrary path connecting any two particles gives the exact difference between the displacement vectors of the two particles.
This guarantees that mean strains on large scales are equal to the spatial average of local values.

Because the area is fixed and the system is densely packed,
changes in density are small.
As a result, we find that the local divergence of the displacement field is negligible.
Most of the deformation takes the form of local shear strains that
are quantified by the curl of the displacement field or vorticity:
$\omega \equiv \partial_y \Delta x - \partial_x \Delta y$, which represents the antisymmetric part of the derivative tensor.
For two dimensional systems the curl is a vector perpendicular to the plane
and thus invariant under rotations within the plane.
Note that it is conventional to associate the magnitude of shear with a symmetric scalar invariant, $\frac{1}{2}\sqrt{(\partial_x \Delta x - \partial_y \Delta y)^2+(\partial_x \Delta y + \partial_y \Delta x)^2}$, the so-called deviatoric strain.
Such symmetric definitions of the strain are completely blind to the vorticity of the displacement which we show below to be quite important in our system.
In particular we will see that the sign of $\omega$ is strongly correlated with the orientation of lines along which shear occurs.
As the magnitude of the strain increases, nonlinear terms in the definition may become important.
The vorticity can be generalized to a rotation tensor $R_{ij}$ defined by the decomposition:
$\delta_{ij} + \partial_i \Delta r_j = S_{ik}R_{kj}$ where $S_{ij}$ is a symmetric tensor which describes pure axial stretching.
In the limit of infinitesimal strains, $\omega$ and the rotation angle associated with $R_{ij}$ become identical.
For finite strains, we find qualitatively similar results for the distribution of $\omega$ and the rotation angle.
 
\section{Spatial Variation of Displacement and Strain}
\begin{figure}
\begin{center}
\includegraphics[width=.84\textwidth]{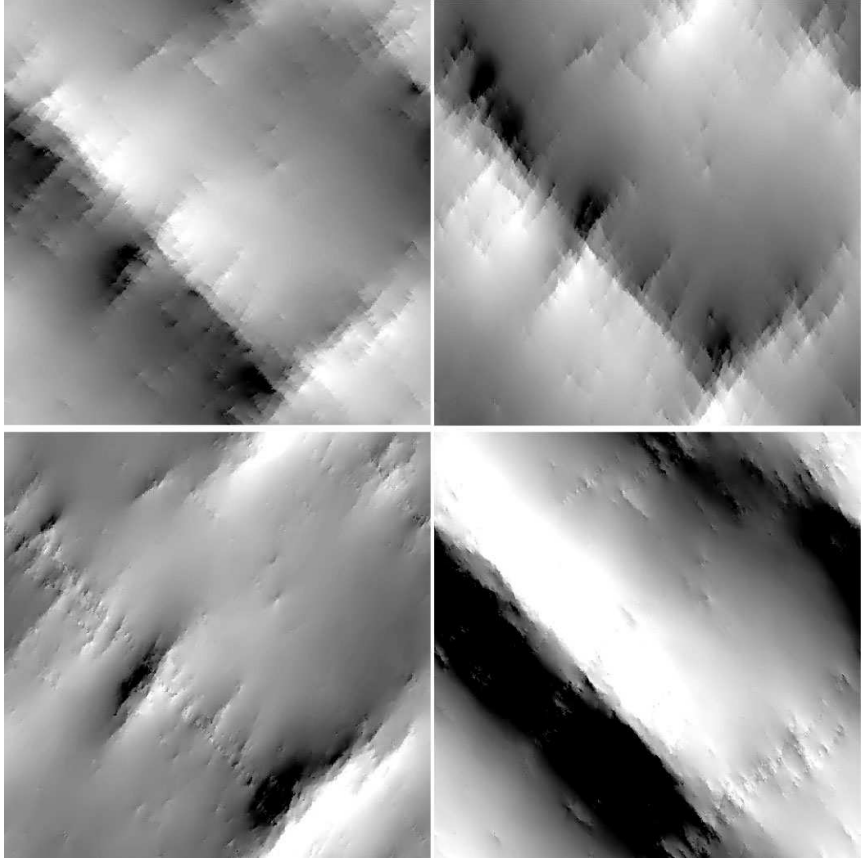}
\end{center}
\caption{
Grayscale plots of the projections of a typical non-affine displacement field $\bf{u}$ onto different directions.
Top left: $\Delta x$.  Top right: $\Delta y$.  Bottom left: $(\Delta x+\Delta y)/\sqrt{2}$.  Bottom right: $(\Delta x-\Delta y)/\sqrt{2}$.
Here $\Delta \gamma=.002$, the x-axis is horizontal, and the gray scale is linear, ranging from $-\sigma_0$ (black) to $+\sigma_0$ (white).
}
\label{fig:dispFieldComponents}
\end{figure}

\begin{figure}
\begin{center}
\includegraphics[width=.5\textwidth]{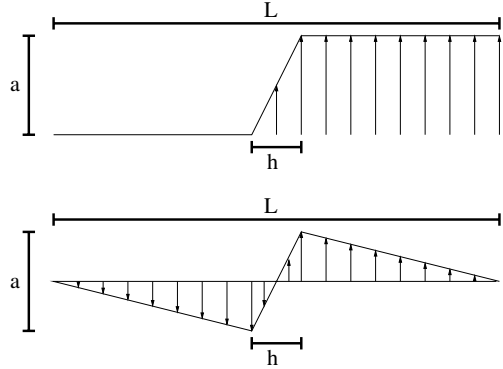}
\end{center}

\caption{
An idealized schematic representation of the (top) total and (bottom) non-affine displacements parallel to a slip line as a function of distance away from the line.
The change $a$ in the total displacement across the periodic cell is accommodated in a plastic zone of width $h$ where the displacement changes rapidly, $\omega \approx a/h$.
The total displacement in the surrounding elastic regions is nearly constant.
Subtracting the affine displacement associated with a strain $\Delta \gamma$ leads to a small gradient, $\omega \approx \Delta \gamma$, in the elastic regions of plots of the non-affine displacement.
}
\label{fig:slipCartoon}
\end{figure}

In figure~\ref{fig:dispFieldComponents}, we plot projections of a typical displacement field for $\Delta \gamma=0.002$ onto horizontal, vertical and diagonal directions:
$\Delta x$, $\Delta y$, $(\Delta x+\Delta y)/\sqrt{2}$, and $(\Delta x-\Delta y)/\sqrt{2}$.
All projections show strong spatial correlations over scales comparable to the
system size, but different projections highlight different features.
For example, the plot for the component proportional to $\Delta {x}-\Delta {y}$ is dominated by a region of strong gradient extended roughly along $y=-x$ while the $\Delta x + \Delta y$ component has a few smaller features extended along $y=x$ and centered roughly about the large feature in $\Delta x - \Delta y$.
The displacements $\Delta x$ and $\Delta y$ show a superposition of all these features.
Note that some of the features extend around the periodic boundary conditions. 

The strong spatial correlations evident in Fig. \ref{fig:dispFieldComponents} show that most of the displacements of atoms that contribute to $P(\Delta r;\Delta \gamma)$ are very far from independent random walks by each particle.
The correlations result from shear along spatially extended slip lines.
The slip lines show up as lines of rapid gradients in Fig. \ref{fig:dispFieldComponents}.
Because they reflect a shear displacement, the amplitude of the displacements is largest for the projection along the slip line.
Thus the slip line that spans the entire system along $\hat{x}-\hat{y}$ shows up most clearly in $\Delta x-\Delta y$, while a number of smaller slip lines along $\hat{x}+\hat{y}$ are emphasized in $\Delta x+\Delta y$.
The reason that slip lines are at an angle of roughly 45$^\circ$ from the compressive axis
is that the stress tensor has the largest shear (off-diagonal)
components for these orientations.
\footnote{Note that the periodic boundary conditions may favor slip line orientations along low order periods such as the diagonals, but we find the angle remains near 45$^\circ$ to the compressive axis even when the cell has deformed significantly.}
Put another way, shear along these lines is most efficient in simultaneously contracting the system along the compressive direction and expanding it in the tensile direction.

Figure \ref{fig:slipCartoon} illustrates (top) the total displacement and (bottom) the non-affine displacement along a slip line as a function of the distance away from the slip line.
The change $a$ in the total displacement across the periodic cell is accommodated almost completely in a plastic zone of width $h$.
The total displacement in the surrounding elastic regions is nearly constant.
Subtracting the affine displacement associated with a strain $\Delta \gamma$ leads to a small gradient $\omega \approx \Delta \gamma = a/L$ in the elastic regions of plots of the non-affine displacement.
For $h << L$ the gradient in the plastic zone is much higher $\omega \approx a/h$.

Traces through Fig. \ref{fig:dispFieldComponents} and similar plots are roughly consistent with Fig. \ref{fig:slipCartoon}.
In all cases there are large regions where the total displacement is nearly constant, leading to a gradual slope $\omega \approx a/L$ in the non-affine displacement.
The amount of strain accommodated in a plastic zone varies somewhat with the strain interval, particularly for slip lines that do not span the system and where there are multiple slip lines in a given strain interval (Fig. \ref{fig:dispFieldComponents} bottom left).
The edge of the plastic zone is also more diffuse than in Fig. \ref{fig:slipCartoon}, leading to a rounding of the peaks in the non-affine displacement.
This rounding varies along the slip line as is evident from the fluctuations along the primary ridge in the bottom right panel of Fig. \ref{fig:dispFieldComponents}.
These variations have important implications for the distributions discussed in the next section.

The form of plots like Fig. \ref{fig:dispFieldComponents}
depends on strain, strain interval and system size.
At very small strains, the entire system responds elastically and the displacements are small.
As the strain increases, the stress and rate of plasticity rise.
For strains of 6\% and greater we find that the system is in a nearly steady state.
In this regime, plastic deformation tends to occur in a series of rapid ``avalanches'' with a wide range of sizes that will be discussed further in future work \cite{SalernoInPrep}.
Between these avalanches the displacement field may be small, but the strain interval between avalanches decreases rapidly with system size.
For the system size considered here, plastic zones that span a large fraction of the system are always observed for $\Delta \gamma \ge 0.002$ and this is the reason we focus on intervals of this order.
Many avalanches occur during this strain interval and they cluster spatially to produce the slip lines evident in Fig. \ref{fig:dispFieldComponents}.

\begin{figure}
\begin{center}
\includegraphics[width=.84\textwidth]{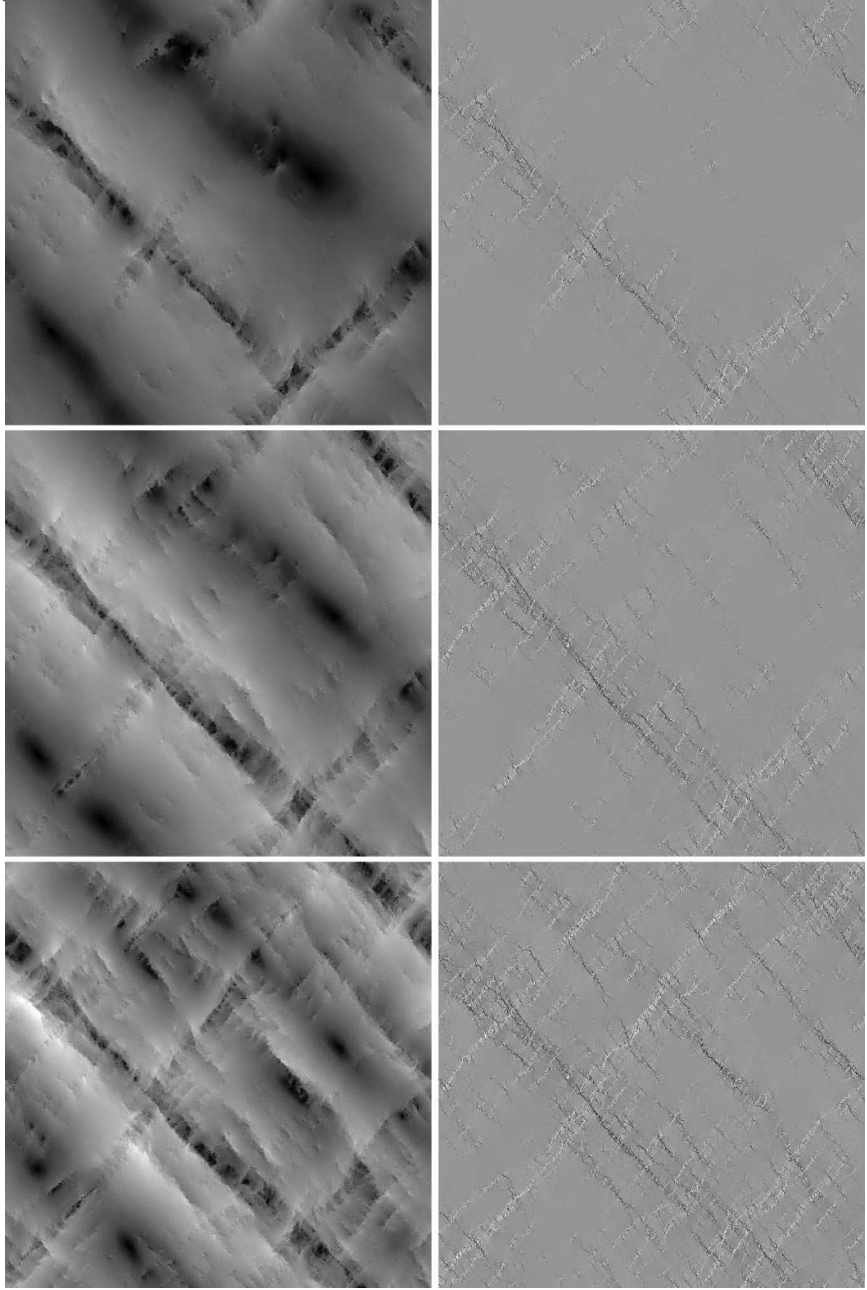}
\end{center}

\caption{
Left: The magnitude, $\Delta r$, of the (non-affine) displacement fields.
Gray scale is linear ranging from $0$ (black) to $+2.5\sigma_0$ (white).
Right: The curl, $\omega$, of the (non-affine) displacement fields.
Gray scale is linear ranging from $-.25$ (black) to $+.25$ (white).
Displacements are computed relative to the same initial state as in Fig. \ref{fig:dispFieldComponents} and over growing strain intervals $\Delta \gamma=.002$ (top), $.004$ (center), and $.008$ (bottom). 
}
\label{fig:dispFieldMags}
\end{figure}

The evolution of the spatial organization of plastically deformed regions is illustrated in Fig. \ref{fig:dispFieldMags}.
The left hand side shows the \emph{magnitude} of the non-affine displacement, $\Delta r$, rather than individual components.
It is these magnitudes that are used to construct the widely studied $P(\Delta r;\Delta \gamma)$.
The right hand side shows $\omega$, which quantifies the amount and direction of local shear.
The successive panels moving downwards in Fig. \ref{fig:dispFieldMags} show the changes over progressively longer strain windows starting from the same initial configuration as in Fig. \ref{fig:dispFieldComponents}.

The families of slip systems along both diagonals contribute equally to $\Delta r$ and are visible in the left hand panels.
However, taking the magnitude of the displacement obscures the nature of the atomic rearrangements.
The slip lines that appeared as sharp steps in $\Delta x+\Delta y$ and $\Delta x - \Delta y$ appear now as regions of \emph{low} $\Delta r$ surrounded by halos of higher $\Delta r$.
The origin of this behavior is evident from Fig. \ref{fig:slipCartoon}.
The peak displacements occur at the outer edges of the plastic zone at a distance of order $h/2$ from the center of the slip line.
The magnitude drops rapidly to zero at the slip line and much more slowly in the elastic region outside.

A better measure of the local plastic response is provided by $\omega$.
The $\omega$ field is clearly localized on the regions of strong shear and much smaller away from these regions.
From Fig.~\ref{fig:slipCartoon} we expect the shear strain to be of order $\Delta \gamma$ in the elastic regions and much higher in the plastic regions that accommodate almost all the deformation of the cell.
The numerical values of $|\omega|$ are consistent with this expectation.
Note that the sign of $\omega$ carries important information about the direction of slip.
In particular, the two families of slip planes produce opposite signs, with a clockwise rotation along lines running parallel to $\hat x - \hat y$ and a counterclockwise rotation on perpendicular slip lines.
Thus plots of $\omega$ highlight the correlations along individual slip lines.

The successive panels moving down Figure \ref{fig:dispFieldMags} show how displacements add as the strain interval increases.
Over short intervals $\Delta \gamma \leq 0.002$ we find that the displacements tend to organize along system-spanning lines.
This allows the entire displacement to be accommodated in plastic zones with minimal energy stored in deformation of elastic regions.
Subsequent slip lines appear to occur at random locations rather than continuing to nucleate along the same path to produce a persistent shear band ({\it e.g.} as in reference \cite{Shi2006qy}).
One way of quantifying this is through the autocorrelation function for displacements over successive intervals.
We find that this correlation function drops rapidly.
By $\Delta \gamma=0.003$ the correlations have dropped by about 2 orders of magnitude and are comparable to the noise.
This is consistent with Fig. \ref{fig:dispFieldMags} where we see a growing number of independent slip lines for $\Delta \gamma \geq 0.004$.
A superposition of randomly located slip lines is consistent with the diffusive growth in $\Delta r$ and $\omega$ discussed in the next section.

The strain interval over which deformations are correlated depends on system size and seems to be related to a characteristic amplitude of the slip across plastic zones.
Examination of many plots like Fig. \ref{fig:dispFieldComponents} shows that $a$ varies from about the mean particle diameter $0.8\sigma_0$ in the central region of shorter slip lines to about twice this value in a system spanning slip line like that in the bottom right of Fig. \ref{fig:dispFieldComponents}.
It is perhaps not surprising that once $a$ is of order a particle size it is energetically favorable to have plastic deformation everywhere along the slip line rather elastic deformation.
Note that the strain interval needed to produce a displacement $a$ is of order $a/L$.
This implies that throughgoing faults should occur over $\Delta \gamma \sim $ 0.001 to 0.002 in our systems.
Our observations are consistent with this and
analysis of the avalanche size distribution also shows that the largest events are spaced by $\Delta \gamma \sim 0.002$.

The plastic regions in Fig. \ref{fig:dispFieldComponents} and on the left of Fig. \ref{fig:dispFieldMags} have a significant width $h \sim 50\sigma$.
Thus the mean shear strain across these regions is only of order $a/h \sim 0.02 - 0.04$ for $\Delta \gamma =0.002$.
While this is an order of magnitude larger than the mean strain $\Delta \gamma$, the peak strains on the right of Fig. \ref{fig:dispFieldMags} are higher still.
Each broad slip zone contains a number of much sharper features with strains that are an order of magnitude higher ($|\omega|=$ 0.2 -0.4).
The width of these smaller features is only a few particle diameters.
More detailed studies of the time dependence show that the slip zones are formed by a large number of avalanches that are spatially correlated to produce the system spanning lines in the figures.
The long-range spatial correlations in the strain field are analyzed in Ref. \cite{MaloneyRobbinsInPrep} and we will focus below on the temporal evolution of the probability distribution of displacement magnitudes and $\omega$.

Comparing results for different system sizes indicates that $h$ increases roughly linearly with $L_0$ while the width of the high strain regions remains of order the particle diameter.
It is interesting to note that the width of fault regions in earthquake systems also tends to scale in rough proportion to the length \cite{Scholz}.
Studies of different system sizes also indicate that the slip amplitude along throughgoing faults is always of order one to two particle diameters.
If $a$ is independent of system size, then the strain interval needed to produce a slip line should be inversely proportional to system size.
There should also be a slower decay with $\Delta \gamma$ in the displacement autocorrelation function.
Results for smaller systems ($L_0=250$ and 500) are consistent with these predictions.

\section{Probability distributions of displacement and strain}

\begin{figure}
\begin{center}
\includegraphics[width=.70\textwidth]{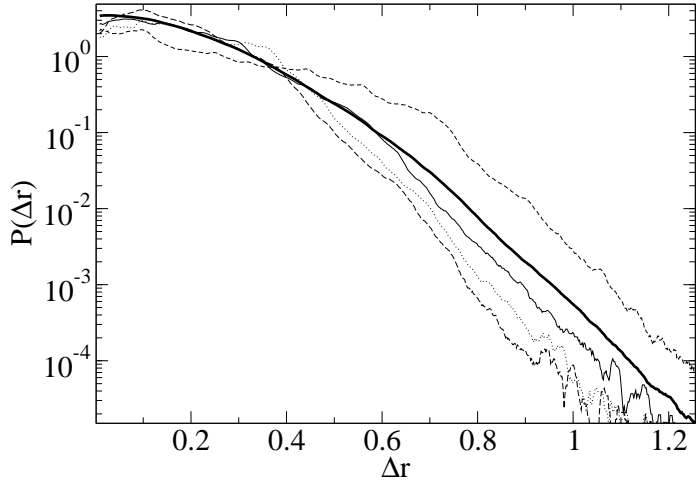}
\end{center}
\caption{$P(\Delta r)$ for several consecutive intervals with $\Delta \gamma=.002$.
The average over all intervals from $\gamma=12$\% to 24\% is shown in heavy black.
}
\label{fig:probDrSeveral}
\end{figure}

Figure \ref{fig:probDrSeveral} shows the probability distribution for finding a given displacement over successive strain intervals with $\Delta \gamma =0.002$.
The distribution $P(\Delta r)$ is normalized so that the integral over all vectors $\bf{\Delta r}$ is unity.
With this definition the distribution goes to a constant as $\Delta r \rightarrow 0$.
Note that the distribution varies significantly from interval to interval.
This reflects the large scale structure of the slip lines discussed above.
In contrast, the distribution of the more local variable $\omega$
is nearly the same for all intervals.
Its average behavior is discussed below.

For an ideal system-spanning slip line with a constant magnitude $a$ and $h<<L$, traces like Fig. \ref{fig:slipCartoon} (bottom) would consist of two straight segments of constant slope $\Delta \gamma$ and $a/h$, respectively.
The distribution $P(\Delta r)$ would be constant from $0$ to $a/2$ and zero for larger displacements.
Given the range of $a$ discussed above, one would expect plateaus extending to $\Delta r \sim 0.4\sigma_0$ to 0.8$\sigma_0$ in Fig. \ref{fig:probDrSeveral}.
The curves show a crossover from gradual to rapid decrease at $\Delta r$ in this range, but do not show true plateaus.
Analysis of displacement fields like those in Figs. \ref{fig:dispFieldComponents} and \ref{fig:dispFieldMags} show that $a$ varies along the slip line.
As discussed above, there is also smearing along the edge of the plastic region that rounds the peaks in traces like Fig. \ref{fig:slipCartoon}.
Interactions between the two families of slip planes also alter the distribution.
Together theses effects lead to a reduction in the range of $\Delta r$ where the distribution for each interval is constant, and an exponential tail in the probability distribution at large magnitudes.
Averaging over all intervals between 12\% and 24\% produces the thick solid line.
The crossover to the exponential tail is broadened by averaging over intervals with different values of $a$ and the exponential tail becomes more pronounced.

\begin{figure}
\begin{center}
\includegraphics[width=.70\textwidth]{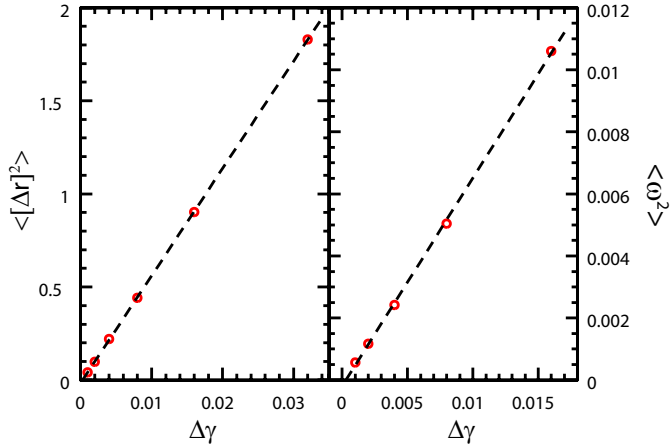}
\end{center}
\caption{Variation with $\Delta \gamma$ of the mean squared (left) displacement $\langle\Delta r^2\rangle$ and (right) strain $\langle\omega^2\rangle$.
Straight-lines are linear fits through the data points, which have statistical errors that are slightly smaller than the symbol size.}
\label{fig:meanSquaredDispAndOmega}
\end{figure}

Figure \ref{fig:meanSquaredDispAndOmega} shows how the mean squared displacement and strain increase with $\Delta \gamma$.
All results are averaged over strains between 12\% and 24\%.
As would be expected for a diffusive process, both $\langle \Delta r^2\rangle$ and $\langle \omega^2 \rangle$ rise linearly with $\Delta \gamma$.
The fit lines have a small negative offset at $\Delta \gamma=0$.
This reflects the correlation in the increase in both quantities at small strains, which cause them to rise quadratically with $\Delta \gamma$ at small strain intervals ($\Delta \gamma \leq 0.001$).
While the linear rise of $\langle \Delta r^2 \rangle$ at large $\Delta \gamma$ is consistent with diffusion, this only implies that the displacements of any given particle in successive strain intervals are uncorrelated.
As shown above, there are strong spatial correlations that imply one should not view the slope of Fig. \ref{fig:meanSquaredDispAndOmega}(left) as a single particle diffusion constant.
A detailed examination of the probability distributions
also reveals some subtle longer-lived correlations in displacement.

\begin{figure}
\begin{center}
\includegraphics[width=.80\textwidth]{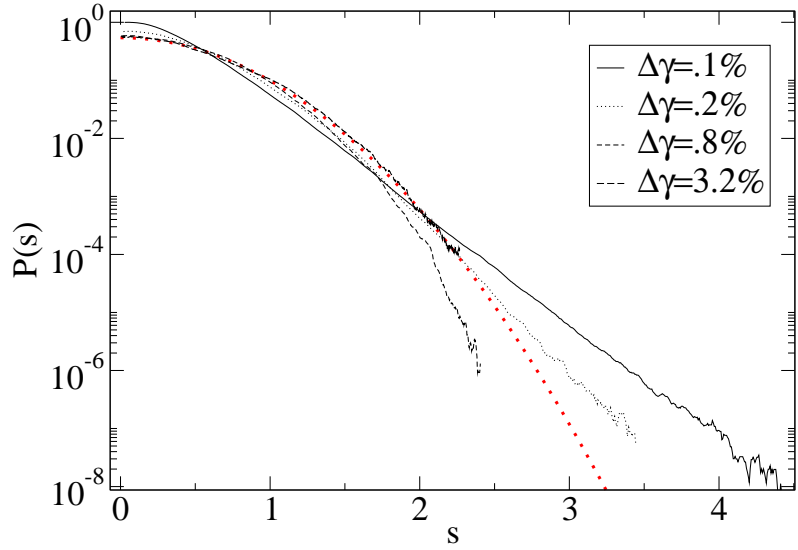}
\end{center}
\caption{Average displacement distributions, $P(\Delta r; \Delta \gamma)$, for $\Delta \gamma=0.001$ (dot-dashed), $0.002$ (dotted),  $0.008$ (dashed), and $0.032$ (solid).
To compensate for the increase in width with $\Delta \gamma$, the curves are plotted as a function of $s \equiv \Delta r / \sigma_0 \left( 100\Delta \gamma \right)^{0.5}$.
The thick (red) dotted line is a one-parameter fit of the $\Delta \gamma=.032$ data to a Gaussian.
Averages are over all intervals from $\gamma=12\%$ to $\gamma=24\%$.
}
\label{fig:probDr}
\end{figure}

Figure~\ref{fig:probDr} shows the probability distribution of displacements averaged over initial states and scaled by $\sqrt{100 \Delta \gamma}$ to remove the mean increase in magnitude shown in Fig.~\ref{fig:meanSquaredDispAndOmega}.
For the smallest interval, the distribution has an exponential tail that extends over eight orders of magnitude.
It is interesting to note that the displacement distributions following individual plastic events or avalanches have also been found to be exponential \cite{Tanguy2006ul,Lemaitre2007ij}.
If successive events were strictly decorrelated, the distributions for longer
intervals would be given by convoluting the distributions for short intervals
and would converge to a Gaussian.
The distributions do seem to become more Gaussian as $\Delta \gamma$
increases, but there is a systematic decrease in the probability of
large events for $\Delta \gamma = 0.008$.
This suggests that the behavior is still correlated, with large displacements
less likely to occur in the same location as in previous intervals.
By $\Delta \gamma =0.032$ these correlations have nearly disappeared and the data are consistent with a Gaussian distribution.
It is interesting to note that at this point the product of the number of throughgoing faults ($\sim 0.032/0.002=16$) and the fraction of the system affected by each ($h/L \sim 50/1000$) is of order one, implying that slip lines must revisit the same regions of space.

\begin{figure}
\begin{center}
\includegraphics[width=.70\textwidth]{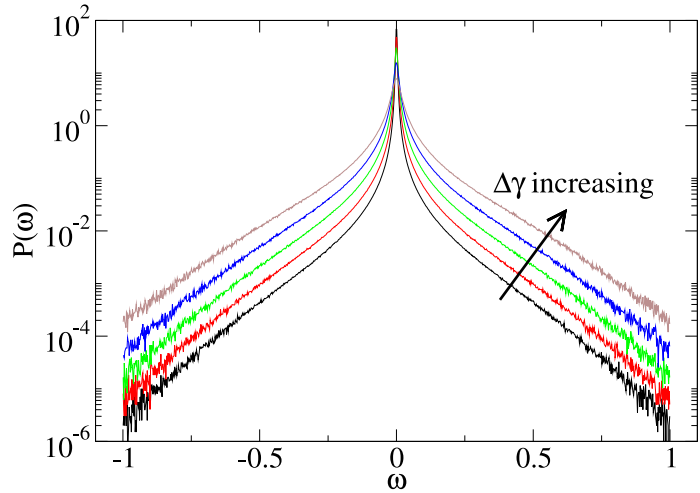}
\end{center}
\caption{The time averaged $\omega$ distributions, $P(\omega; \Delta \gamma)$ for $\Delta \gamma=.001,.002,.004,.008,.016$.}
\label{fig:probOmega}
\end{figure}

Figure~\ref{fig:probOmega} shows the time averaged $\omega$ distributions for various $\Delta \gamma$.
The distributions consist of central peaks that cross over to exponential tails.
The exponential tails
persist even at the largest $\Delta \gamma$.
Roughly speaking, regions of the material with small strains away from any shear zones (the gray regions in the right column of Figure~\ref{fig:dispFieldMags}) contribute to the "elastic" central peaks, while the strongly sheared material in the core of the shear zones (the black and white regions in the right column of Figure~\ref{fig:dispFieldMags}) contributes to the "plastic" exponential tails.
To first order, weight is removed from the central peak and redistributed to the tails as the amount of plastic strain increases.

\begin{figure}
\begin{center}
\includegraphics[width=.70\textwidth]{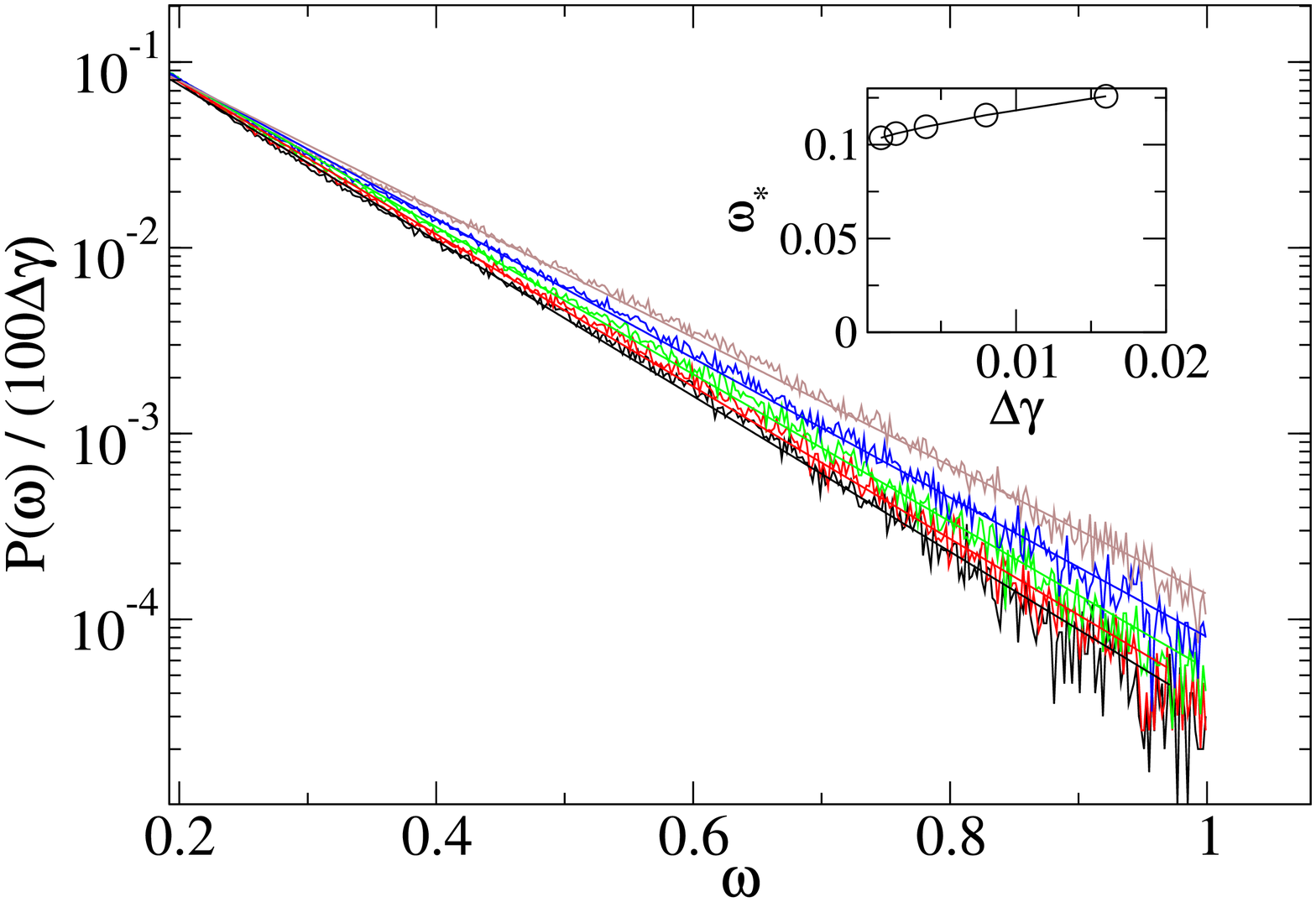}
\end{center}
\caption{Distribution of $\omega$ scaled by $\Delta \gamma$ in percent
for $\Delta \gamma=.001,.002,.004,.008,.016$.
Curves increase with $\Delta \gamma$ at large $\omega$.
Straight lines show exponential fits, $P\propto e^{-\omega/\omega_*}$, and the inset shows $\omega_*$ as a function of $\Delta \gamma$.
}
\label{fig:probOmegaExpFit}
\end{figure}

To test this picture quantitatively, the $\omega$ distributions are scaled by $\Delta \gamma$ and replotted in figure~\ref{fig:probOmegaExpFit}.
This rescaling comes close to collapsing the data in the tails for the smallest $\Delta \gamma$, but becomes less satisfactory for increasing $\Delta \gamma$.
Direct convolution of data from small intervals shows similar deviations from probabilities for larger intervals.
This implies that there is some additional correlation over the range of $\Delta \gamma$ considered here.
Fits to an exponential decay, $P(\omega) \sim \exp(\omega/\omega^*)$, show a slight increase in $\omega^*$ with $\Delta \gamma$ (inset).
This increase implies that increasing $\Delta \gamma$ does not merely increase the size of the plastic region and thus the total weight in the tails.
In addition, there is a slight increase in the relative weight of large magnitude regions that must reflect a correlated increase in strain in regions of large strain.

It is interesting to note that the value of $\omega^*$ has a natural physical interpretation.
The vorticity $\omega =\partial_y \Delta x - \partial_x \Delta_y$ while the shear strain is $0.5*[\partial_y \Delta x + \partial_x \Delta_y]$.
Thus $\omega^*=0.1$ corresponds to a shear strain of about 5\%, which is close to the yield strain \cite{Rottler2005uq}
As noted above, the regions of large $\omega$ have strains much bigger than the mean strain in the plastic region ($a/h$) but small enough that there are several in the plastic zone that combine to produce the total displacement.
The probability distribution suggests that the yield strain sets the characteristic scale for these high strain regions.

Another important parameter is the strain interval required to produce plastic deformation throughout the entire system. 
One can estimate this from the integral under the exponential tails in $P(\omega)$ at small strains, and the assumption that the weight grows approximately linearly with $\Delta \gamma$ as in Figure \ref{fig:probOmegaExpFit}.
The result is that the regions of plastic strain should cover the system after a strain of about 10\%.
This is larger than the estimate of 3\% strain required for plastic zones to cover the system, but as noted above, only a fraction of the plastic zones has the large strains indicative of local plasticity.

\section{Discussion}

Although ours is, to our knowledge, the first study of strain distributions in simulations of sheared systems, several studies have looked at displacement distributions in athermal systems.
All considered simple shear, $v_x = \dot{\gamma} y$, rather than the pure shear strain studied here.
Some simulations used quasi-static simulations with small strain steps~\cite{Tanguy2006ul,Lacks2002lr,Lemaitre2007ij}, while others used a constant slow shear rate as in our simulations~\cite{Ono2003fk,Ono2002qy,Radjai2002qy}.
The quasi-static studies all employed Lennard-Jones potentials and imposed simple shear with
Lees-Edwards periodic boundary conditions~\cite{Lacks2002lr,Lemaitre2007ij} or rigid walls parallel to the flow direction~\cite{Tanguy2006ul}.
Ono and co-workers~\cite{Ono2003fk,Ono2002qy} simulated steady shear of wet foams using Durian's bubble model~\cite{DurianPhysRevE551739}, while Radjai and Roux simulated granular media with dissipation entering through friction at particle contacts \cite{Radjai2002qy}.
Only the latter work and our simulations include inertia.
Despite the wide range of protocols and interactions, almost all studies find diffusive behavior for the non-affine displacement.
\footnote{
For simple shear this is most easily measured in the displacement \emph{perpendicular} to the flow direction $x$,
$\langle \Delta y^2 \rangle\propto \Delta \gamma$.
}
The only exception is the work of Radjai and Roux~\cite{Radjai2002qy} who found $\langle \Delta y^2 \rangle\propto \Delta \gamma^{1.8}$.
It seems very important to determine the cause of this striking difference.
The only unique feature of these simulations appears to be the incorporation of friction forces, which are not generally thought to lead to such qualitative changes in behavior.

The value of the effective diffusion constant $D_{eff}\equiv \langle \Delta r ^2 \rangle /\Delta \gamma$ can be estimated from our observations of slip lines and the simple model in Fig. \ref{fig:slipCartoon}.
This displacement field gives a constant distribution of displacements from $-a/2$ to $a/2$, implying a mean-squared displacement of $a^2/12$.
If system-spanning slip lines produce independent displacements at strain intervals of $a/L$, then
$\langle \Delta r^2 \rangle \approx (\Delta \gamma /(a/L)) a^2/12$
or $D_{eff} = La/12$.
Using the value of $D_{eff} = 57 \sigma_0^2$ from Fig. \ref{fig:meanSquaredDispAndOmega},
one finds $a \approx 0.7 \sigma_0$.
Improving the estimate of the mean-squared displacement from each slip line leads to larger
values of $a$ since the actual distribution of displacements 
(Fig. \ref{fig:probDrSeveral}) is cut off before $a/2$.
We conclude that the measured diffusion is consistent with the direct observation that slip lines have amplitudes of one to two particle diameters and contribute independently to the displacement.

This model for the effective diffusion constant implies a linear scaling with system size.
We have not yet examined this scaling directly, but recent studies of
a similar model by Lemaitre and Caroli~\cite{Lemaitre2007ij} showed a pronounced increase in $D_{eff}$ with system size.
Data for the two largest systems, $L=20$ and 40, appear consistent with linear scaling with $L$.
Results for $L=10$ are larger than expected, but there may be significant finite size effects in these systems.
Note that we find $h/L \sim 1/20$ so system sizes with $L < 20$ could not have the plastic zone occupying the same fraction of the system.
Extrapolating Lemaitre and Caroli's result
\footnote{Note that in reference~\cite{Lemaitre2007ij}, the imposed homogeneous \emph{simple} shear flow field is defined as $v_x=\dot{\gamma}y$, $v_y=0$.
The associated symmetric strain rate tensor associated with this flow field has a magnitude of $\dot{\gamma}/2$, so one needs to scale the values of $\langle \Delta y^2 \rangle / \Delta \gamma$ from reference~\cite{Lemaitre2007ij} by a factor of $2$ to make a direct comparison to our results for \emph{pure} shear geometry, $v_x=\dot{\gamma}x, v_y=-\dot{\gamma}y$.},
for $L=40$, $D_{eff} = (2.2 \pm .1) \sigma_0^2 $ to $L=1000$ yields $D_{eff} \approx (55 \pm 2.5) \sigma_0^2$, which is consistent to our measured value of $57 \sigma_0^2$.
The magnitude of $\Delta \gamma$ needed to observe diffusive behavior is also roughly a factor of 25 smaller in our systems.
These results provide a dramatic check of the linear scaling with system size in diffusion, although we caution that these comparisons should only be viewed qualitatively, as the impact of different shearing geometries has not been taken into account.  

Lemaitre and Caroli discuss the system-size dependence of their results in terms of displacements produced by independent avalanches rather than independent slip lines.
Earlier studies of individual plastic events showed
that the rate of large avalanches scales with the system length~\cite{Maloney2004dk,Maloney2006dz}.
Based on the assumption that the total strain $\Delta \gamma$ was accommodated by these avalanches, Maloney and Lemaitre concluded that the displacement $a_a$ associated with each avalanche was independent of system size and of order $0.1 \sigma_0$.
Extrapolating their results~\cite{Maloney2004dk,Maloney2006dz} for $L \sim 50$ to our system sizes would give large avalanches spaced by $\Delta \gamma=0.0002$, an order of magnitude more frequent than the rate of system-spanning slip lines in our simulations.
This is consistent with our conclusion that the
slip lines in Figs. 1 and 3 are not the result of individual avalanches, but the correlated displacement produced by multiple avalanches. 
Assuming that each avalanche contributes independently to the displacement would give too small a value of $D_{eff}$.

System size dependence of $D_{eff}$ could be crucial in analyses which invoke the notion of an effective temperature based on an effective Stokes-Einstein relationship between $D_{eff}$ and the viscosity $\eta$ \cite{Ono2002qy,Lacks2002lr}.
The product of diffusion and viscosity is related to the temperature in equilibrium systems.
In systems with a yield stress, the viscosity diverges like the inverse strain rate, $\eta \sim \Delta t/\Delta \gamma$, at low strain rates where the stress reaches a plateau.
As we've seen here, one also expects the effective diffusion constant to grow linearly with the strain rate: $\Delta r^2/ \Delta t = (\Delta r^2 / \Delta \gamma) (\Delta\gamma/\Delta t)$.
This means that the product of the viscosity and diffusivity should become constant at small enough strain rate: $(\Delta r^2 / \Delta t) \eta \sim \Delta r^2 / \Delta \gamma$.
However our results imply that this constant should have a dramatic size dependence, and one must proceed with caution to utilize it as an effective temperature.

The scaling of $D_{eff}$ would be very different if the correlations in displacement did not span the system.
There is some evidence that system spanning slip lines are suppressed by increasing temperature \cite{PhysRevE68011507,Varnik2004qy}.
They may also be suppressed by the surrounding fluid in experiments on colloidal systems \cite{Stevens}.
Dissipation through flow in the fluid is reduced by maintaining a linear flow profile that may inhibit localization of the plastic deformation.
Studies of both effects would be of interest.

The most important message in our results is that one must go beyond the second moments of the displacement and strain distributions to get insight into the microscopic processes at play in sheared systems.
Displacement distributions may be nearly Gaussian, scaling with $\sqrt{\Delta t}$, while at the same time, there may be strong spatial correlations and heterogeneities in the system.
The strain distributions can remain \emph{far} from Gaussian, even while the displacement distributions become nearly Gaussian.
Such behavior provides a clear signature of spatial organization.
Furthermore, while individual particle trajectories may seem jerky with smooth motion interrupted by abrupt jumps, these jumps should not necessarily be interpreted as a particle escaping the cage formed by its neighbors (as is often assumed to be the case), but could just as easily correspond to coherent displacement of the neighborhood with very little \emph{relative} motion of particles.
In the future, it will prove interesting to repeat the analysis here focusing on strain in addition to the more commonly studied displacement in thermal ({\it e.g.} super-cooled liquids) or brownian({\it e.g.} colloidal suspensions)  systems on approach to the jamming transition.
This kind of measurement could serve as an important probe of the co-operative nature of the dynamics.

\ack
This material is based on work supported by the National Science Foundation under Grant No. DMR-0454947, CTS-0320907 and PHY-99-07949.


\end{document}